\documentclass[twoside]{article}
\usepackage[latin1]{inputenc}
\usepackage{t1enc}
\usepackage{a4}
\usepackage{tabularx}
\usepackage{epsf}

\def\bref{\vspace{4pt}\noindent\hangindent=10mm}

\def\spose#1{\hbox to 0pt{#1\hss}}
\def\simlt{\mathrel{\spose{\lower 3pt\hbox{$\mathchar"218$}}
     \raise 2.0pt\hbox{$\mathchar"13C$}}}
\def\simgt{\mathrel{\spose{\lower 3pt\hbox{$\mathchar"218$}}
     \raise 2.0pt\hbox{$\mathchar"13E$}}}

\def\etal{{\rm et~al. }}
\def\hi{H\,{\sc i}~}
\def\hii{H\,{\sc ii}~}

\begin{document}

\setcounter{figure}{0}
\setcounter{section}{0}
\setcounter{equation}{0}

\begin{center}
{\Large\bf
Continuous star formation in gas-rich\\[0.2cm]
dwarf galaxies}\\[0.7cm]

Simone Recchi, Gerhard Hensler\\[0.17cm]
Institute of Astronomy, Vienna University\\
T\"urkenschanzstrasse 17, A-1180 Vienna, Austria \\
recchi@astro.univie.ac.at, hensler@astro.univie.ac.at
\end{center}

\vspace{0.5cm}

\begin{abstract}
\noindent{\it
  Blue Compact Dwarf and Dwarf Irregular galaxies are generally
  believed to be unevolved objects, due to their blue colors, compact
  appearance and large gas fractions.  Many of these objects show an
  ongoing intense burst of star formation or have experienced it in
  the recent past.  By means of 2-D hydrodynamical simulations,
  coupled with detailed chemical yields originating from SNeII, SNeIa
  and intermediate-mass stars, we study the dynamical and chemical
  evolution of model galaxies with structural parameters similar to
  IZw18 and NGC1569.  Bursts of star formation with short duration are
  not able to account for the chemical and morphological properties of
  these galaxies.  The best way to reproduce the chemical composition
  of these objects is by assuming long-lasting episodes of star
  formation and a more recent burst, separated from the previous
  episodes by a short quiescent period.  The last burst of star
  formation, in most of the explored cases, does not affect the
  chemical composition of the galaxy observable in \hii regions, since
  the enriched gas produced by young stars is in a too hot phase to be
  detectable with the optical spectroscopy.  }
\end{abstract}

\section{Introduction}

Among dwarf galaxies, Blue Compact Dwarfs (BCDs) and Dwarf Irregulars
(dIrrs) are characterized by large gas content and often an active
star formation (SF).  They also show very blue colors and low
metallicities and are therefore commonly believed to be poorly evolved
systems.  They are consequently ideal targets to study the feedback
between star formation and interstellar medium.  They have also been
suggested to be the local counterparts of faint blue objects detected
in excess at z $\sim$ 1 (Babul \& Rees 1992; Lilly et al.  1995).

It has recently become clear that most of these objects show the
presence of stars of intermediate-old age (Kunth \& \"Ostlin 2000),
but their importance for the global metallicity and energy budget of
the galaxy is still unknown.  It is interesting to simulate galaxies
whose light and colors are dominated by young stars (like IZw18 and
NGC1569) and to see whether their chemical and morphological
properties are dominated by a recent burst of star formation or
whether older episodes of SF are required in order to explain some of
their characteristics.

In general, the SF in BCDs is described as a {\it bursting} process
(Searle et al. 1973), namely, short, intense episodes of SF are
separated by long inactivity periods.  A {\it gasping} mode of SF
(long episodes of SF of moderate intensity separated by short
quiescent periods) is instead often used to describe the star
formation in dIrrs (Aparicio \& Gallart 1995).  Good galaxy candidates
experiencing gasping star formation are for instance NGC6822 (Marconi
et al. 1995), Sextans B (Tosi et al. 1991) and the LMC (Gallagher et
al. 1996).  These two different SF regimes have been tested in the
framework of chemical evolution models (Bradamante et al. 1998;
Chiappini et al. 2003a; Romano et al. 2004), producing similar
results, therefore is it not easy to discriminate between these two
different SF scenarios on the basis of chemical evolution models
alone.  Our aim is to simulate the dynamical and chemical evolution of
model galaxies by means of a 2-D hydrodynamical code in cylindrical
coordinates, coupled with detailed chemical yields.  Since the largest
set of parameters is derived from various observations for IZw18 and
NGC1569, we intend to see whether is it possible to put constraints on
their past SF history.

Despite the different classification (IZw18 is a BCD galaxy, whereas
NGC1569 is often classified as dIrr), these two objects show similar
properties: both of these objects are in the aftermath of an intense
burst of SF, are very metal poor (0.02 Z$_\odot$ for IZw18, Izotov \&
Thuan 1999; 0.23 Z$_\odot$ for NGC1569, Gonz\'alez Delgado et al.
1997) and have an extremely large gas content.

In spite of their simplicity, the chemical composition of gas-rich
dwarf galaxies is often peculiar and hardly understandable in terms of
closed-box models.  In particular, the N/O ratios are puzzling.  For
metallicities larger than 12 + log (O/H) $\sim$ 7.8, the log (N/O) is
linearly increasing with the metallicity, although with a large
scatter.  This is consistent with a secondary production of nitrogen
(N synthesized from the original C and O present in the star at
birth).  At lower metallicities, all the galaxies seem to show a
constant log (N/O) (of the order of -1.55/-1.6), with almost no
scatter (Izotov \& Thuan 1999; hereafter IT99).  This is a typical
behaviour of element produced in a primary way (starting from the C
and O newly formed in the star).  Although some specific metal-poor
objects seem to contradict the existence of the plateau (see e.g.
Skillman et al. 2003; Pustilnik et al. 2004), an explanation is needed
in order to understand the behavior of most dwarf galaxies in this
range of metallicities.

This problem has been investigated in several papers and various ideas
have been proposed to solve this puzzle.  Izotov \& Thuan (1999)
proposed a significant primary production of nitrogen in massive
stars, whereas the models of Henry et al. (2000) were able to explain
the low log (N/O) at low metallicities with a very weak and constant
star formation rate.  Recently, K\"oppen and Hensler (2004) proposed
the infall of metal-poor gas as a mechanism able to reduce the oxygen
abundance of the galaxy, keeping the N/O ratio constant.  For none of
these models, however, a complete investigation of the structural and
energetic effects by means of hydrodynamical simulations have been
performed.

Moreover, IZw18 and NGC1569 have been carefully studied in the past
and now we know, with reasonable accuracy, the chemical abundances not
only in the \hii medium, but also in other gas phases.  In particular,
recent {\sl FUSE} (Far Ultraviolet Spectroscopic Explorer) data
allowed two groups of astronomers (Aloisi \etal 2003; Lecavelier des
Etangs \etal 2004) to calculate the \hi abundances in IZw18.  There is
also an attempt to evaluate the chemical composition of the hot medium
in the galactic wind of NGC1569, (Martin, Kobulnicky \& Heckman 2002).
It is challenging to compare the results of our simulations with these
observations, in order to see whether is possible to put additional
constraints on the past SF activity of IZw18 and NGC1569.

\section{Chemical and dynamical evolution of IZw18}

We first describe the evolution of a model galaxy resembling IZw18.
This object, the most metal-poor galaxy locally known, has been
considered in the past as a truly ``young'' galaxy, experiencing star
formation for the very first time, since old stars could not be
observed.  Moreover, the spectral energy distribution is well
described assuming a single, recent burst of SF (Mas-Hesse \& Kunth
1999; Takeuchi et al.  2003).  An underlying old population of stars
has been first observed by Aloisi et al. (1999) in the optical and
\"Ostlin (2000) in the infrared, the age of which being however still
disputed.  This age ranges from some hundred Myrs (Aloisi et al.
1999) to a few Gyrs (\"Ostlin 2000).

In the attempt of reproducing the characteristics of IZw18, we
consider both a bursting and a gasping SF scenario.  We first assume a
couple of instantaneous bursts, separated by a quiescent period of
300-500 Myr.  In the second subsection, we will consider a gasping SF
with an old episode of SF lasting 270 Myr at a SF rate of 6 $\times$
10$^{-3}$ M$_\odot$ yr$^{-1}$, a gap of 10 Myr and a recent burst
lasting only 5 Myr and being 5 times more intense then the
long-lasting episode.  This SF history has been suggested by Aloisi et
al. (1999) by fitting the observed Color-Magnitude diagram of IZw18
with synthetic ones.  In this attempt at reproducing the main
characteristics of IZw18, we also vary the slope of the IMF and the
adopted nucleosynthetic yields, both for massive and for
intermediate-mass stars (IMS).  The model parameters are summarized in
Table 1.

We use a 2-D hydrodynamical code with source terms.  The input of
energy and chemical elements into the systems is provided by SNeII
(mainly responsible for the production of $\alpha$-elements), SNeIa
(source of most of the iron-peak elements) and winds from low- and
intermediate-mass stars (responsible for the bulk of nitrogen
production and for a significant fraction of carbon).  In all the
considered models we will assume that SNeIa are more effective than
SNeII in thermalizing the interstellar medium.  This is due to the
fact that SNeIa explode in a warmer and more diluted medium, owing to
the previous activity of SNeII.  The details about the code can be
found in Recchi et al. (2001; 2002).

\subsection{Bursting mode of star formation}

The first (instantaneous) episode of SF produces 10$^5$ M$_\odot$ of
stars and is separated from the second one by a quiescent period of
300-500 Myr.  We follow the evolution of the ISM after the first
burst.  After the assigned inactivity interval, we calculate how much
cold gas is remained in the central part of the galaxy and we convert
10\% of this gas into stars.  The metallicity of this new stellar
population is given by the metallicity of the gas which the starts are
formed from.  We obtain a second burst of SF with a mass of $\sim$ 5
$\times$ 10$^5$ M$_\odot$ and a metallicity of 1/50 Z$_\odot$.

The first burst is not able to account for the metallicity of the gas
in IZw18.  At the onset of the second burst, there is a sudden
increase of the oxygen content (and, consequently, a sudden decrease
of C/O and N/O abundance ratios).  This mode of SF is therefore
characterized by huge variations of the chemical composition of the
galaxy on very short timescales.  A few tens Myrs after the onset of
the second burst, N/O and C/O abundance ratios begin to grow, due to
the release of chemical elements from IMS.  The results of our
simulations match the abundance ratios found in the literature for two
age intervals of the last burst: between 4 and 7 Myr and between
$\sim$ 40 and $\sim$ 80 Myr.  The second solution does not fit neither
the morphology nor the spectral energy distribution of IZw18 and has
to be rejected.  The favoured age of the last burst is then, in the
framework of a bursting scenario of SF, between 4 and 7 Myr.  This
solution has a very short duration, since the oxygen abundance is
increasing very rapidly in this phase.  Models with a SF gap of 300 or
500 Myr show approximately the same behaviour.

\subsection{Gasping mode of star formation}

As we have seen in the previous section, the bursting mode of SF can
fit the observed abundances and abundance ratios found in literature
for IZw18 only in tiny intervals of time.  This can also be seen in
Fig. 1, where we have plotted the evolution of a bursting model with
300 Myr of inactivity between two instantaneous bursts (model IZw --
1; long-dashed line).  This model crosses the values of log (N/O) and
log (O/H) inferred by IT99, but, as stated in the previous section,
the N/O abundance ratio in particular shows large variations on short
time-scales and remains within the allowed range of values only very
briefly.  In this section we therefore study what happens if we relax
the hypothesis of instantaneous bursts of SF.  The way in which
continuous episodes of SF can be treated by our code is described in
Recchi et al. (2004).

\begin{table}[ht]
\begin{flushleft}
\caption{Parameters for the IZw18 models}
\end{flushleft}
\label{model}
\begin{flushleft}
\begin{tabular}{ccccc}
  \hline\hline
\noalign{\smallskip}

  Model  &  SF mode$^a$ & x (IMF slope) & IMS yields & 
Massive stars yields\\
\noalign{\smallskip}

  \hline 
  IZw -- 1 & bursting & 1.35 & RV81$^b$ & WW95$^c$\\
  IZw -- 2 & gasping  & 1.10 & VG97$^d$ & WW95    \\
  IZw -- 3 & gasping  & 1.35 & MM02$^e$ & MM02    \\
  IZw -- 4 & gasping  & 1.35 & VG97     & WW95    \\
  \hline
 \end{tabular}
\end{flushleft}
$^a$ Bursting (2 instantaneous bursts of SF) or gasping (long episode 
of SF plus a recent burst), as described in Sect. 2.

$^b$ Renzini \& Voli (1981)

$^c$ Woosley \& Weaver (1995)

$^d$ van den Hoek \& Groenewegen (1997)

$^e$ Meynet \& Maeder (2002)
\end{table}

Models with Salpeter IMF, yields from massive stars coming from
Woosley \& Weaver (1995) and IMS yields coming from the most cited
papers (van den Hoek \& Groenewegen 1997; Renzini \& Voli 1981)
overestimate the nitrogen content of the galaxy by 0.4 -- 0.6 dex and
some of them underestimate O/H (see Recchi et al. 2004).  The best fit
between the observations and the results of the model is obtained when
implementing the yields of both massive and IMS from Meynet \& Maeder
(2002).  In their models, nitrogen is mainly produced in a primary way
through rotational diffusion of carbon in the hydrogen-burning shell.
This set of nucleosynthetic yields has given good results in chemical
evolution models (Chiappini et al. 2003b) and are therefore worth
testing in our simulations.  Is it, however, necessary to point out
that these models do not take into consideration the last phases of
the stellar evolution (in particular the third dredge-up) and may
therefore underestimate the total amount of nitrogen produced in IMS.

The results of this model are shown in Fig. 1 (model IZw -- 2; dotted
line).  This model nicely reproduces the log (N/O) of IZw18 and
slightly overestimates the observed oxygen content.  Another
possibility to get results closer to the observations is by using a
flatter IMF (with a slope of x $= 1.1$) in order to produce more
oxygen in massive stars.  This model is also shown in Fig.  1 (model
IZw -- 3; short-dashed line).  The log (N/O) is $\sim$ 0.2 dex larger
than the observations, therefore it better reproduces the observations
compared with models adopting Salpeter IMF and van den Hoek \&
Groenewegen (1997) IMS yields.

\begin{figure}[ht]
 \epsfxsize=11.5cm \epsfbox{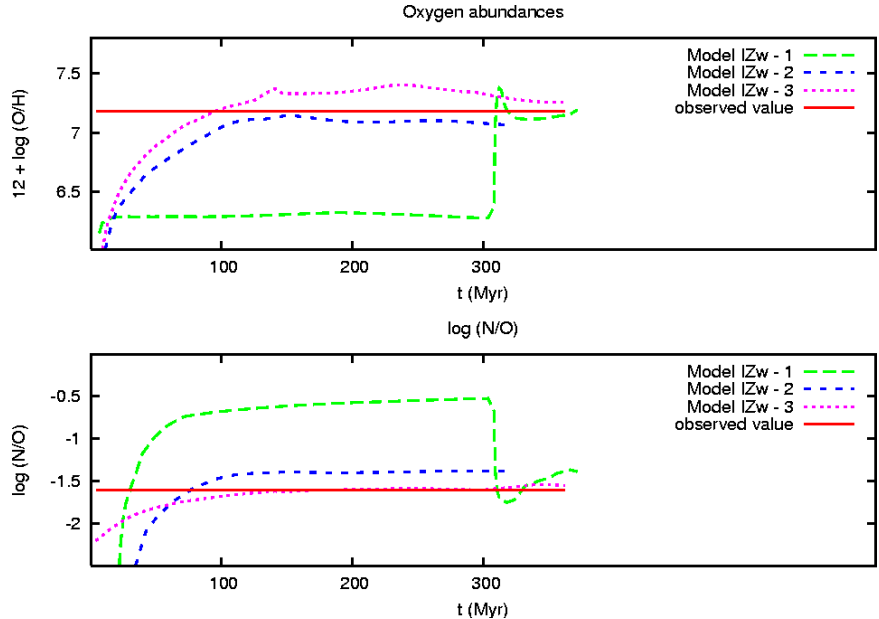}
  \caption{
    Evolution of 12 + log (O/H) (upper panel) and log (N/O) (lower
    panel) for IZw18 models.  The solid line represents the observed
    values found for IZw18 (IT99).  Model parameters are summarized in
    Table 1.}
\end{figure}

It is also worth noticing in Fig. 1 that, after the onset of the
second burst of SF (i.e. at 280 Myr), the metallicity of the galaxy
for the gasping models (IZw -- 2 and IZw -- 3) does not show any
sudden increase, at variance with what happens in the framework of a
bursting scenario of SF (model IZw -- 1; long-dashed line).  This is
due to the fact that the first SF episode is sufficiently energetic to
create an outflow.  The metals produced by the second generation of
stars are released in a hot medium or directly channelled along the
outflow.  Consequently, they do not have the chance to cool down to
temperatures detectable with the optical spectroscopy.  This means
that, if the SF in a galaxy has been active long enough and has been
energetic enough, the last burst of SF does not affect at all the
metallicity of the gas. The so-called {\it self-pollution} of galaxies
by freshly produced metals (Kunth \& Sargent 1986) can hold only under
particular conditions, i.e. for the very first bursts of SF or if the
gap between subsequent episodes is long enough.

Another way to see the different chemical evolution of bursting and
gasping models is by plotting the log (N/O) vs. log (O/H) diagram
(Fig. 2).  The solid line is the evolution of the IZw -- 3 model,
whereas the behaviour of the IZw -- 4 model is shown as dashed line.
The bursting model IZw -- 1 (dotted line) is plotted starting from the
onset of the last burst.  In Fig.  2 are also plotted the most recent
data available in literature about O and N abundances in very
metal-poor galaxies.  As stated in the introduction, according to the
results of IT99, these galaxies form a ``plateau'' in this diagram,
the (N/O) ratios being very similar at different metallicities.
Indeed, isolating the data of IT99 (filled squares and filled
triangle), this plateau is pretty evident, whereas, when adding
observations coming from other authors, the scatter seems to increase.
We probably need more statistics and better measurements before
drawing firm conclusions.

\begin{figure}[ht]
\begin{center}
\epsfxsize=8cm \epsfbox{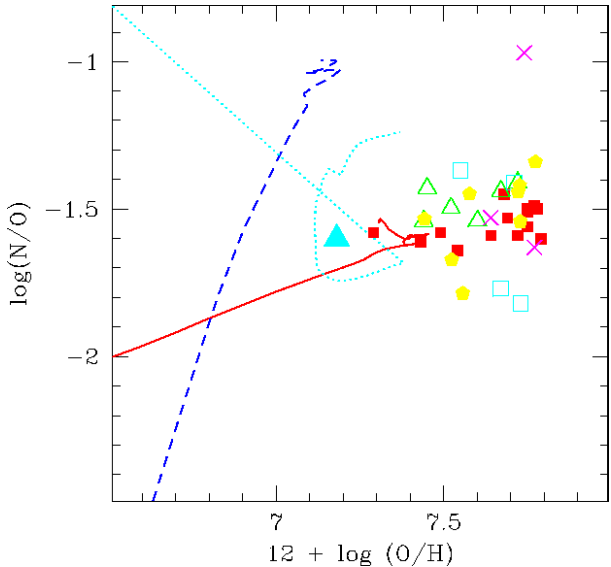}
  \caption[recchifig1]{
    log (N/O) vs. 12 + log (O/H) in metal-poor BCD galaxies.  The
    filled triangle is the IZw18 value calculated by IT99.  Filled
    squares are other galaxies measured by IT99.  The collection of
    data by van Zee et al. (1997) is shown in open squares.  The
    measurements by Kobulnicky \& Skillman (1996) are indicated by
    filled pentagons.  Crosses represent the values tabulated by
    V\'ilchez \& Iglesias-P\'aramo (2003).  The other data points
    (open triangles) are taken from different sources.  Also shown is
    the evolution in the N/O vs. O/H plane of two gasping models:
    model IZw -- 3 (solid line) and model IZw -- 4 (dashed line).  The
    dotted line represent the evolution of model IZw -- 1 (bursting
    model), after the onset of the second burst of SF.  Model
    parameters are summarized in Table 1}
\end{center}
\end{figure}

Model IZw -- 4 exceeds the nitrogen of the most metal-poor galaxies
(in particular the nitrogen of IZw18), whereas model IZw -- 3 (solid
line) matches the low log (N/O).  It is also worth noticing that for
this model it takes $\sim$ 100 Myr to reach the O abundance of the
most metal-poor galaxies (see Fig. 1).  After that, the chemical
evolution tracks span a tiny region of the diagram for the remaining
$\sim$ 200 Myr of the evolution of the galaxy.  The bursting model IZw
-- 1 (dotted line) shows instead large abundance variations in very
short time-scales.  It takes only $\sim$ 80 Myr for this model to
reach the final point.  This kind of SF regime would produce a large
scatter in the log (N/O) vs.  log (O/H) diagram even at low
metallicities.  Under the hypothesis of a gasping SF regime the
abundance ratios are stable for long time-scales, justifying the lack
of scatter and the apparent plateau in the log (N/O) vs. log (O/H)
diagram at low metallicities.


Two recent papers (Aloisi et al. 2003; Lecavelier des Etangs et al.
2004) tried to derive the chemical composition of the \hi medium of
IZw18.  Even starting from the same {\sl FUSE} data, the two papers
differ significantly in the final abundance determinations.  In
particular, Lecavelier des Etangs et al. (2004) found an oxygen
abundance in the neutral medium similar to the O/H derived in the \hii
regions, whereas Aloisi et al. (2003) obtained an O abundance a factor
$\sim$ 3--4 lower than in the ionized gas.  Due to the larger oxygen,
the N/O ratio calculated by Lecavelier des Etangs et al.  (2004) is
much below the observations in the \hii regions.  The determination of
Aloisi et al. (2003) is instead similar to the N/O found the in the
ionized phase.  In our models, we find a slight underabundance of O in
the neutral medium, but not enough to justify the observations of
Aloisi et al. (2003).  The calculated log (N/O) is instead more
consistent with the determinations of Lecavelier des Etangs et al.
(2004) (see Recchi et al. 2004).  Due to the differences in the
results of these two groups, no robust constraints can be imposed by
means of this comparison.  A more careful parametrical study of
gasping models for IZw18 will be presented in a forthcoming paper
(Recchi et al. 2005, in prep.)

\section{Chemical and dynamical evolution of NGC1569}

We adopt the same hydrodynamical code described in the previous
section to model a galaxy resembling NGC1569, a prototypical starburst
galaxy.  This galaxy is particularly valuable as a case study due to
its proximity (2.2 Mpc according to Israel (1988); 1.95 or 2.8 Mpc
according to Makarova \& Karachentsev (2003)).  It consists of two
super star clusters (SSCs), with an absolute separation of 80--85 pc.
The IMF slope of the SSCs is well constrained by the luminosity/mass
ratio and is close to the Salpeter slope (Sternberg 1998).  The
stellar population in NGC1569 is dominated by stars younger than a few
tens Myrs, the majority of which are found in two prominent super star
clusters (Anders et al. 2004), although the presence of older stars
have been inferred (Vallenari \& Bomans 1996; Greggio et al. 1998).

We therefore adopted two possible SF histories for NGC1569.  The first
is a single burst of star formation, lasting for 25 Myr.  The SF rate
inferred by Greggio et al. (1998), assuming a Salpeter IMF, is 0.5
M$_\odot$ yr$^{-1}$.  A weaker SF rate for the present burst (0.13
M$_\odot$ yr$^{-1}$) has been found in more recent studies of the CMD
diagram of NGC1569 (Angeretti et al. 2005).  Martin et al.  (2002), by
fitting the H$\alpha$ luminosity, found a SF rate of 0.16 M$_\odot$
yr$^{-1}$ for the last burst.  Given the uncertainties of this value,
we keep the SF rate as a free parameter.

A more complex episode of SF has been recently discovered by Angeretti
et al.  (2005) and is characterized by 3 episodes of SF.  The first
happened between 600 and 300 Myr ago at a rate of 0.05 M$_\odot$
yr$^{-1}$.  This episode is followed by a period of inactivity of 150
Myr and then by a second episode lasting 110 Myr at a SF rate of 0.04
M$_\odot$ yr$^{-1}$.  After a short quiescent period (3 Myr) the last
episode of SF started.  The onset of this episode is therefore 37 Myr
ago, lasting until 13 Myr ago (24 Myr of duration in total, consistent
with the estimates of Anders et al.  2004) at a rate of 0.13 M$_\odot$
yr$^{-1}$.  It is worth pointing out that, in this work, the stars in
a field of 200 $\times$ 200 pc have been analyzed and, therefore, the
gaps in the SF process might be spurious.  Table 2 summarizes the
parameters adopted to model NGC1569.

\begin{table*}[ht]
\caption{Parameters for the NGC1569 models}
\label{model}
\begin{flushleft}
\begin{tabular}{cccc}
  \hline\hline
\noalign{\smallskip}

  Model  &  SF episodes & SF rate (M$_\odot$ yr$^{-1}$) 
& M$_{gas}$ (M$_\odot$) \\
\noalign{\smallskip}

  \hline 
  NGC -- 1 & 1 & 0.1              &             10$^8$ \\
  NGC -- 2 & 1 & 0.5              &             10$^8$ \\
  NGC -- 3 & 3 & 0.05; 0.04; 0.13 &             10$^8$ \\
  NGC -- 4 & 3 & 0.05; 0.04; 0.13 & 1.8 $\times$ 10$^8$ \\
  \hline
 \end{tabular}
\end{flushleft}
\end{table*}

\vspace{1cm}
\subsection{Single episode of star formation}

In the models described in this section, the SF lasts for 25 Myr and
the SF rate is a free parameter.  We have simulated two model
galaxies: one with a SF rate of 0.1 M$_\odot$ yr$^{-1}$ (model NGC --
1) and one with a rate of 0.5 M$_\odot$ yr$^{-1}$ (model NGC -- 2; see
Table 2).  It is hard to fine-tune the SF rate: a large rate (model
NGC -- 1) injects a large amount of energy into the ISM.  This energy
drives a very powerful galactic wind, able to push away from the
galaxy most of the pristine gas at variance with what is observed in
NGC1569.  On the other hand, in the model with a lower SF rate (model
NGC -- 2), the oxygen produced by massive stars does not reach the
observed abundance of NGC1569 of 12 + log (O/H) = 8.19 (Kobulnicky \&
Skillman 1997). In Fig. 3 we plot the evolution of oxygen and N/O for
the model NGC -- 1.  As we can see, this model is neither able to
explain the amount of oxygen present in the galaxy nor the (N/O)
abundance ratio.

\begin{figure}[ht]
  \epsfxsize=11.5cm \epsfbox{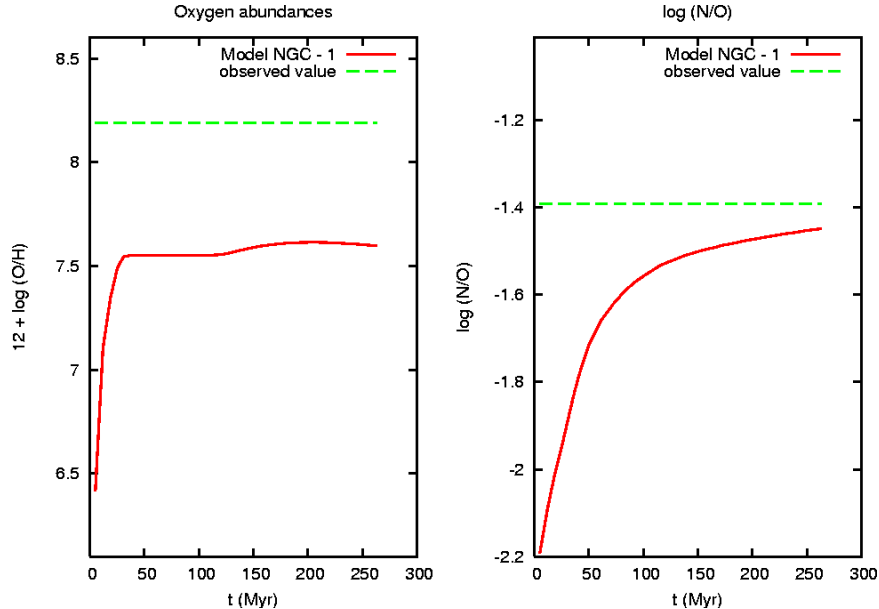}
  \caption{
    Evolution of 12 + log (O/H) (left panel) and log (N/O) (right
    panel) for a NGC1569 model with a single episode of SF at a rate
    of 0.1 M$_\odot$ yr$^{-1}$ (model NGC -- 1; solid lines).  The
    dashed lines are the observed values found by Kobulnicky \&
    Skillman (1997).}
\end{figure}

As anticipated in the introduction, Martin et al. (2002) were able to
give an estimate of the metallicity of the hot X-ray emitting gas in
the galactic wind.  This information completes the puzzle of
understanding the metal enrichment.  Even if the single-burst models
are not able to account for the chemical and morphological properties
of NGC1569, it is none the less interesting to calculate the
metallicity of the hot gas (i.e. of the gas with temperatures larger
than 0.3 keV) and compare it with the estimates of Martin et al.
(2002).  This comparison is shown in Fig.  4 for the NGC -- 2 model.
At the moment of the onset of a galactic wind, the oxygen abundance of
the hot phase is already 1/4 of solar.  It increases up to 2 times
solar after $\sim$ 50 Myr from the beginning of the burst.  The arrows
drawn in the plot represent the estimates of the oxygen content of the
galactic wind of NGC1569 (best fit; upper and lower limits).  The
oxygen composition of the hot gas increases continuously in this phase
since, after 50 Myr, massive stars are still exploding and releasing
oxygen into the interstellar medium.  At later times however, the
oxygen composition begins to decrease due to the larger fraction of
pristine gas ablated from the supershell and entrained in the galactic
wind.  The [O/Fe] ratio is initially larger than solar due to the fact
that the break-out occurs when SNeIa are not yet releasing their
energy and metals into the ISM.  At later times, since the SNeIa eject
their products from the galaxy very easily (Recchi et al. 2001), the
[O/Fe] ratio decreases.  The observations of Martin et al. (2002)
point toward a galactic wind dominated by $\alpha$-elements, therefore
the outflow is probably still triggered by SNeII.

\begin{figure}[ht]
 \epsfxsize=10cm \epsfbox{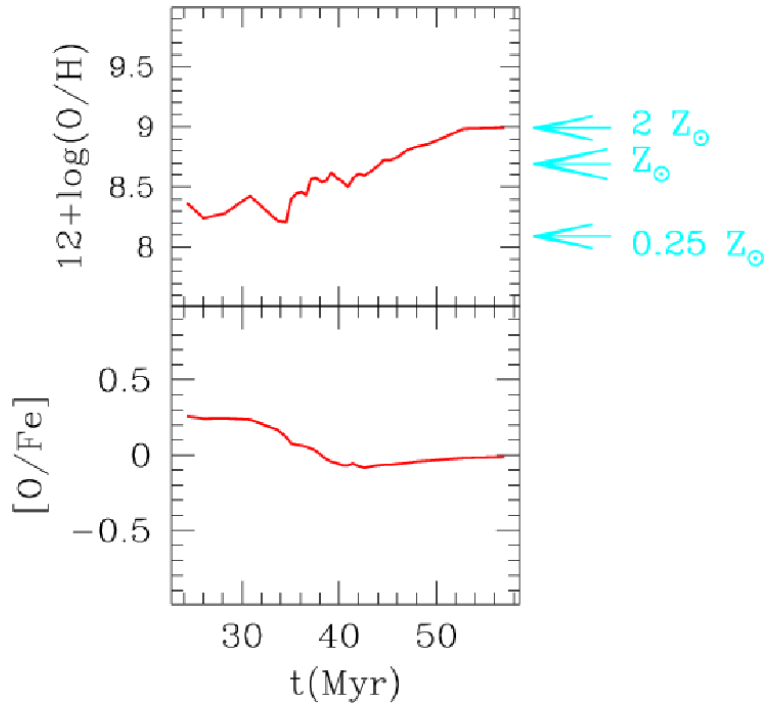}
  \caption{
    Evolution of 12 + log (O/H) (upper panel panel) and [O/Fe]
    abundance ratio (lower panel) for the hot gas entrained in the
    galactic wind for the model NGC -- 2 (see Table 2).  The arrows
    indicate the oxygen abundance of the galactic wind inferred by
    Martin et al. (2002).}
\end{figure}

\subsection{Three episodes of star formation}

As shown in the previous section, single SF bursts of short durations
are not able to account for the global properties of NGC1569.  We
therefore describe in this section the evolution of models in which
the SF is a gasping process, occurring in 3 different episodes as
explained in Sect. 3.  Since in this case the SF rate is no longer a
free parameter, we decide to explore the effect of a different initial
ISM distribution.  In particular, we consider a ``light'' model (model
NGC -- 3), in which the total galactic \hi mass at the beginning of
the simulation is $\sim$ 10$^8$ M$_\odot$ and a model with a factor of
2 more gas initially present inside the galaxy (model NGC -- 4; see
Table 2).  Since there are still uncertainties about the total gas
mass of NGC1569 (see e.g. Stil \& Israel 2002; M\"uhle et al.  2003)
one has the freedom to test different values of the initial mass and
to see which one is more appropriate to reproduce the characteristics
of NGC1569.  We adopt hereafter the nucleosynthetic prescriptions of
Meynet \& Maeder (2002), since they seem to give the better
description of the chemical properties of IZw18 (see Sect. 2), bearing
in mind that the predicted nitrogen can be a lower limit.

In Fig. 5 we show the evolution of oxygen (upper panel) and log (N/O)
(lower panel) for the models NGC -- 3 (dashed lines) and NGC -- 4
(dotted lines) and we compare them with the abundances derived from
Kobulnicky \& Skillman (1997) (solid lines).  At the end of the
simulations (after $\sim$ 600 Myr), the oxygen is reproduced nicely by
the model NGC -- 4 and also model NGC -- 3 is close to the observed
value.  It is worth noticing that the outflow created by the
pressurized gas is very weak in the NGC -- 4 model and of moderate
intensity for the model NGC -- 3.  The fraction of oxygen lost through
the galactic wind is larger in the light model.  In the first hundreds
of Myrs the oxygen abundance predicted by the model NGC -- 3 is
larger, since it is diluted by a smaller amount of hydrogen.  At later
times, however, the O abundance predicted by this model slightly
decreases with time, since some oxygen is expelled from the galaxy.

\begin{figure}[ht]
 \epsfxsize=11.5cm \epsfbox{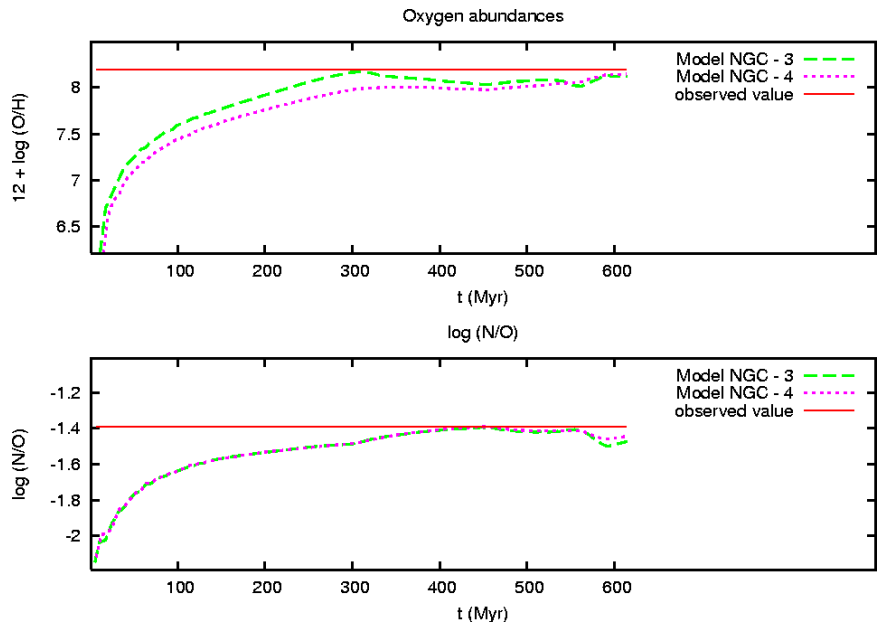}
  \caption{
    Evolution of 12 + log (O/H) (upper panel) and log (N/O) (lower
    panel) for two NGC1569 models in which the Angeretti et al. (2005)
    SF history is implemented.  The solid line represents the observed
    values found for NGC1569 (Kobulnicky \& Skillman 1997).  The
    dashed line is the evolution of model NGC -- 3, whereas the dotted
    line shows the evolution of model NGC -- 4 (see Table 2).}
\end{figure}

The final log (N/O) predicted by these models are $-$ 1.47 (model NGC
-- 3) and $-$ 1.44 (model NGC -- 4).  These values slightly
underestimate the observations of Kobulnicky \& Skillman (1997), but
are still reasonably close to it, considering the observational errors
(0.05 dex).

\subsection{Model with a big infalling cloud}

\begin{figure}[ht]
 \epsfxsize=11.5cm \epsfbox{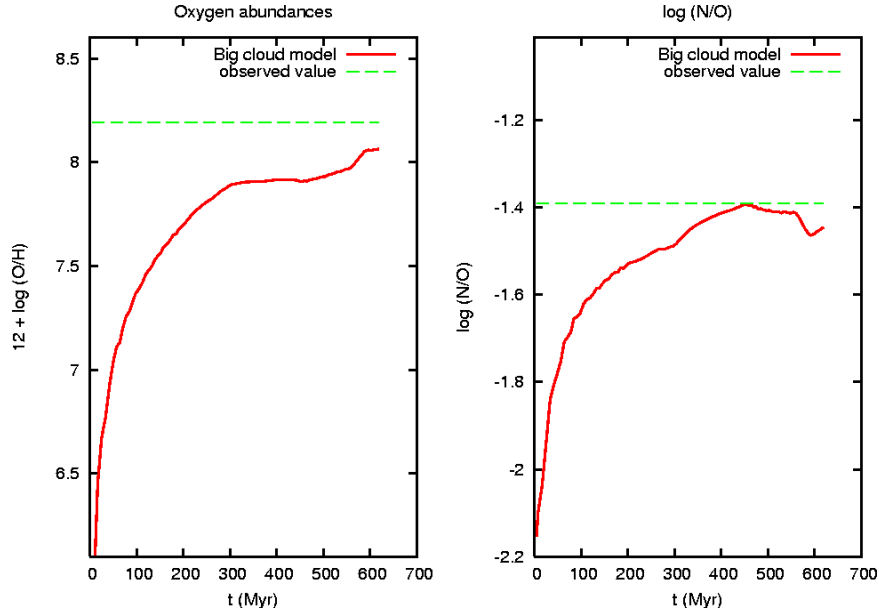}
  \caption{
    Evolution of 12 + log (O/H) (left panel) and log (N/O) (right
    panel) for a NGC1569 models in which the infall of a big cloud
    towards the center of the galaxy is taken into consideration.}
\end{figure}

Observations show the presence of extended \hi clouds and complexes
surrounding NGC1569 (Stil \& Israel 1998).  In particular, there is a
series of gas clumps in the southern halo probably connected with a
\hi arm present in the western side of the galaxy.  These \hi features
can be attributed to the debris of a tidally disrupted big cloud
infalling towards NGC1569 (M\"uhle et al. 2005).  The mass of this
complex is difficult to asses.  The lower limit given by M\"uhle et
al. (2005) is 1.2 $\times$ 10$^7$ M$_\odot$ (the sum of the mass of
all the detected groups of clouds), but some \hi could have been
already accreted.  If this complex is similar to the high velocity
clouds in the Local Group, one has to expect masses larger than a few
10$^7$ M$_\odot$ (Blitz et al. 1999).  As a first attempt to study the
effect of a big cloud infalling towards the galaxy, we assume a mass
of 2 $\times$ 10$^7$ M$_\odot$.  The initial position of the cloud is
2 kpc away from the center of the galaxy along the polar axis (due to
the assumed symmetry of the system, this is the only reasonable
initial configuration).  The infalling velocity of this cloud is 10 km
s$^{-1}$, similar to the local sound speed, and its radius is 1 kpc.
The other structural parameters are as in model NGC -- 3 (see Table
2).

The evolution of oxygen and log (N/O) for this model is shown in Fig.
6.  The development of the galactic wind is hampered by the pressure
of this big infalling cloud and, consequently, no major outflow is
developed during the simulation.  The oxygen abundance is therefore
always increasing, since only a negligible fraction of it is lost from
the galaxy.  The final log (N/O) is consistent with the observations,
whereas this model underestimates the final oxygen content of the
galaxy (by $\sim$ 0.15 dex).  This is due to the fact that the big
cloud on its path towards the galaxy sweeps up and drags some gas
initially present in the outer regions of the galaxy.  The final gas
mass inside the galaxy is therefore larger than the initial one.  It
is worth noticing that the prominent outflow visible in NGC1569
(Martin et al. 2002) deviates from the results of this simulation.
Therefore, we either have to consider a larger input of energy into
the system or have to consider a different infall direction of the
cloud.  Indeed, observations show that this \hi complex seems to wrap
around the disk of NGC1569 and to approach the galaxy from the western
side.  A more careful parametrical study of model galaxies reproducing
NGC1569 will be presented in a forthcoming paper (Recchi \& Hensler
2005, in prep.)

\section{Conclusions}

By means of a 2-D hydrodynamical code, we have studied the chemical
and dynamical evolution of model galaxies resembling IZw18 and
NGC1569, two gas-rich dwarf galaxies in the aftermath of an intense
burst of SF.  We have considered in both cases either episodes of SF
of short duration (bursting SF), or more complex SF behaviours, in
which the galaxies have experienced in the past long-lasting episodes
of SF, separated from the last more intense burst by short periods of
inactivity (gasping star formation).

Models with a bursting star formation are generally unable to account
for the chemical and morphological properties of these two objects.
In the case of IZw18, they produce huge variations of the chemical
composition of the galaxy in short time-scales.  They are able to fit
at the same time the C, N, O composition of IZw18, but only for very
short time intervals.  This pattern of the chemical tracks would
presumably give rise to a large scatter in the abundance ratios.  The
observations available nowadays disagree with this scenario, since
most metal-poor galaxies seem to share the same [N/O] abundance ratio
(IT99).  In the case of NGC1569, models with a single short episodes
of SF either severely underproduce O or inject too much energy into
the system, enough to unbind a too large fraction of the gas initially
present in the galaxy.

The best way to reproduce the chemical composition of both, IZw18 and
NGC1569, is therefore assuming long-lasting, continuous episodes of SF
of some hundreds Myrs of age and a recent and more intense short
burst.  Adopting the star formation prescriptions derived from the
comparison of the color-magnitude diagrams with synthetic ones (Aloisi
et al. 1999 for IZw18; Angeretti et al. 2005 for NGC1569) we produce
results in good agreement with the observations, if the yields of
Meynet \& Maeder (2002) are implemented.

For what concerns NGC1569, a model in which a big cloud is falling
towards the center of the galaxy along the polar axis inhibits almost
completely the formation of a galactic wind, at variance with what
observed.

In most models with gasping star formation, the presently observed
chemical composition of the galaxy reflects mostly the chemical
enrichment from old stellar populations.  In fact, if the first
episodes of SF are powerful enough to create a galactic wind or to
heat up a large fraction of the gas surrounding the star forming
region, the metals produced by the last burst of star formation are
released in a too hot medium or are directly expelled from the galaxy
through the wind.  They do not have the chance to pollute the
surrounding medium and contribute to the chemical enrichment of the
galaxy.

\subsection*{Acknowledgments}

We thank Stefanie M\"uhle and Luca Angeretti for interesting
discussions.  S.R. acknowledges generous financial support from the
Alexander von Humboldt Foundation and Deutsche Forschungsgemeinschaft
(DFG) under grant HE 1487/28-1.  The Observatory of Vienna is also
acknowledged for the travel support.

\subsection*{References}

{\small
\bref 
Aloisi, A., Savaglio, S., Heckman, T.M., Hoopes, C.G.,
Leitherer, C., Sembach, K.R. 2003, ApJ, 595, 760

\bref
Aloisi, A., Tosi, M., Greggio, L. 1999, AJ, 118, 302

\bref
Anders, P., de Grijs, R., Fritze-v. Alvensleben, U., 
Bissantz, N. 2004, MNRAS, 347, 17

\bref
Angeretti, L., Tosi, M., Greggio, L., Sabbi, E., Aloisi, A., 
Leitherer, C. 2005, AJ, submitted

\bref
Aparicio, A., Gallart, C. 1995, AJ, 110, 2105


\bref 
Babul, A., Rees, M.J. 1992, MNRAS, 255, 346

\bref
Blitz, L., Spergel, D.N., Teuben, P.J., Hartmann, D., Burton, W.B. 
1999, ApJ, 514, 818

\bref
Bradamante, F., Matteucci, F., D'Ercole, A. 1998, A\&A, 337, 338

\bref
Chiappini, C., Matteucci, F., Meynet, G. 2003b, A\&A, 410, 257

\bref
Chiappini, C., Romano, D., Matteucci, F. 2003a, MNRAS, 339, 63

\bref
Gallagher, J.S. et al. 1996, ApJ, 466, 732

\bref
Gonz\'alez Delgado, R.M., Leitherer, C., Heckman, T., Cervi\~no, M. 
1997, ApJ, 483, 705

\bref
Greggio, L., Tosi, M., Clampin, M., de Marchi, G., Leitherer, 
C., Nota, A., Sirianni, M. 1998, ApJ, 504, 725

\bref
Henry, R.B.C., Edmunds, M.G., K\"oppen, J. 2000, ApJ, 541, 660

\bref
Israel, F.P. 1988, A\&A, 194, 24

\bref
Izotov, Y.I., Thuan, T.X. 1999, ApJ, 511, 639 (IT99)

\bref
Kobulnicky, H.A., Skillman, E.D. 1996, ApJ, 471, 211

\bref
Kobulnicky, H.A., Skillman, E.D. 1997, ApJ, 489, 636

\bref
K\"oppen, J., Hensler, G. 2004, A\&A, submitted

\bref
Kunth, D., \"Ostlin, G. 2000, A\&AR, 10, 1

\bref
Kunth, D., Sargent, W.L.W. 1986, ApJ, 300, 496

\bref 
Lecavelier des Etangs, A., D\'esert, J.-M., Kunth, D.,  
Vidal-Madjar,A., Callejo, G., Ferlet, R., H\'ebrard, G., 
Lebouteiller, V. 2004, A\&A, 413, 131

\bref 
Lilly, S.J., Tresse, L., Hammer, F., Crampton, D., Le Fevre, O. 
1995, ApJ, 455, 108

\bref
Makarova, L.N., Karachentsev, I.D. 2003, Ap, 46, 144

\bref
Marconi, G., Tosi, M., Greggio, L., Focardi, P. 1995, AJ, 109, 173

\bref
Martin, C.L., Kobulnicky, H.A., Heckman, T.M. 2002, ApJ, 
574, 663

\bref
Mas-Hesse, J.M., Kunth, D. 1999, A\&A, 349, 765

\bref
Meynet, G., Maeder, A. 2002, A\&A, 390, 561

\bref
M\"uhle, S., Klein, U., Wilcots, E.M., H\"uttenmeister, S. 
2003, ANS, 324, 40

\bref
M\"uhle, S., Klein, U., Wilcots, E.M., H\"uttenmeister, S. 
2005, submitted

\bref
\"Ostlin, G. 2000, ApJ, 535, L99

\bref
Pustilnik, S., Kniazev, A., Pramskij, A., Izotov, Y., Folz, C., 
Brosch, N., Martin, J.-M., Ugryumov, A. 2004, A\&A, 419, 469

\bref
Recchi, S., Matteucci, F., D'Ercole, A. 2001, MNRAS, 322, 800

\bref
Recchi, S., Matteucci, F., D'Ercole, A., Tosi, M. 2002, 
A\&A, 384, 799

\bref
Recchi, S., Matteucci, F., D'Ercole, A., Tosi, M. 2004, 
A\&A, 426, 37

\bref
Renzini, A., Voli, M. 1981, A\&A, 94, 175

\bref
Romano, D., Tosi, M., Matteucci, F. 2004, to appear in the proceedings of 
the conference ``Starbursts - From 30 Doradus to Lyman break galaxies'', 
eds. R. de Grijs and R.M. Gonzalez Delgado (Kluwer)

\bref 
Searle, L., Sargent, W.L.W., Bagnuolo, W.G. 1973, ApJ, 179, 427

\bref
Skillman, E.D., C\^ot\'e, S., Miller, B.W. 2003, AJ, 125, 610

\bref
Sternberg, A. 1998, ApJ, 506, 721

\bref
Stil, J.M., Israel, F.P. 1998, A\&A, 337, 64

\bref
Stil, J.M., Israel, F.P. 2002, A\&A, 392, 473

\bref
Takeuchi, T.T., Hirashita, H., Ishii, T.T., Hunt, L.K., Ferrara, A. 2003, 
MNRAS, 343, 839

\bref
Tosi, M., Greggio, L., Marconi, G., Focardi, P. 1991, AJ, 102, 951

\bref
Vallenari, A., Bomans, D.J. 1996, A\&A, 313, 713

\bref
van den Hoek, L.B., Groenewegen, M.A.T. 1997, A\&AS, 123, 
305

\bref
van Zee, L., Haynes, M.P., Salzer, J.J. 1997, AJ, 114, 2479

\bref
V\'ilchez, J.M., Iglesias-P\'aramo, J., 2003, ApJS, 145, 225

\bref
Woosley, S.E., Weaver, T.A. 1995, ApJS, 101, 181
}

\vfill

\end{document}